\documentclass[traditabstract]{aa} 
\usepackage{txfonts}
\usepackage{graphicx}

\begin{document}

\title{Binaries discovered by the SPY survey}
\subtitle{
VI. Discovery of a low mass companion to the hot subluminous planetary nebula central star EGB\,5 - A recently ejected common envelope?
\thanks{Based on observations at the Paranal Observatory of the European Southern Observatory for programmes No. 167.H-0407(A) and 71.D-0383(A). Based on observations collected at the Centro Astron\'omico Hispano Alem\'an (CAHA) at Calar Alto, operated jointly by the Max-Planck Institut f\"ur Astronomie and the Instituto de Astrof\'isica de Andaluc\'ia (CSIC). Some of the data used in this work were obtained at the William Herschel Telescope (WHT) operated by the Isaac Newton Group of Telescopes (ING).}
}

\author{S. Geier \inst{1}
   \and R. Napiwotzki \inst{2}
   \and U. Heber \inst{1} 
   \and G. Nelemans \inst{3}}

\offprints{S.\,Geier,\\ \email{geier@sternwarte.uni-erlangen.de}}

\institute{ Dr. Karl Remeis-Observatory \& ECAP, Astronomical Institute,
Friedrich-Alexander University Erlangen-Nuremberg, Sternwartstr. 7, D 96049 Bamberg, Germany
   \and Centre of Astrophysics Research, University of Hertfordshire, College Lane, Hatfield AL10 9AB, UK
   \and Department of Astrophysics, Radboud University Nijmegen, P.O. Box 9010, NL-6500 GL Nijmegen, The Netherlands}

\date{Received \ Accepted}

\abstract{Hot subdwarf B stars (sdBs) in close binary systems are assumed to be formed via common envelope ejection. According to theoretical models, the amount of energy and angular momentum deposited in the common envelope scales with the mass of the companion. That low mass companions near or below the core hydrogen-burning limit are able to trigger the ejection of this envelope is well known. The currently known systems have very short periods $\simeq0.1-0.3\,{\rm d}$. Here we report the discovery of a low mass companion ($M_{\rm 2}>0.14\,M_{\rm \odot}$) orbiting the sdB star and central star of a planetary nebula EGB\,5 with an orbital period of $16.5\,{\rm d}$ at a minimum separation of $23\,R_{\rm \odot}$. Its long period is only just consistent with the energy balance prescription of the common envelope. The marked difference between the short and long period systems will provide strong 
constraints on the common envelope phase, in particular if the masses of the sdB stars can be measured accurately. Due to selection effects, the fraction of sdBs with low mass companions and similar or longer periods may be quite high. Low mass stellar and substellar companions may therefore play a significant role for the still unclear formation of hot subdwarf stars. Furthermore, the nebula around EGB\,5 may be the remnant of the ejected common envelope making this binary a unique system to study this short und poorly understood phase of binary evolution. 

}

\maketitle

\section{Introduction \label{sec:intro}}

The ESO SN Ia Progenitor Survey (SPY) was developed to identify double degenerate progenitor candidates to SN\,Ia. More than $1000$ white dwarfs (WDs) and pre-WDs were checked for radial velocity (RV) variations using high resolution spectra obtained with the UVES instrument at the ESO-VLT (e.g. Napiwotzki et al. \cite{napiwotzki03}). Results for nine binaries discovered in the SPY survey were  presented in papers I-V (Napiwotzki et al. \cite{napiwotzki01}, \cite{napiwotzki02}; Karl et al. \cite{karl03}; Nelemans et al. \cite{nelemans05}; Geier et al. \cite{geier10a}).

Subluminous B stars, which are also known as hot sudwarf stars, display the same spectral characteristics as main-sequence stars of spectral type B, but are much less luminous. They are assumed to be core helium-burning stars with very thin hydrogen envelopes and masses around $0.5\,M_{\rm \odot}$ (Heber \cite{heber86}). The formation of these objects is still unclear. Different formation channels have been discussed (see Han et al. \cite{han02}, \cite{han03}). It is found that a large fraction of sdB stars are members of short period binaries (Maxted et. al \cite{maxted01}; Napiwotzki et al. \cite{napiwotzki04a}). For these systems, common envelope (CE) ejection is the most probable formation channel (Paczy\'nski \cite{paczynski76}). In this scenario, two main-sequence stars of different masses evolve in a binary system. The heavier one will reach the red giant phase first and fill its Roche lobe. If the mass transfer to the companion is dynamically unstable, a common envelope is formed. Owing to gra\-vitational drag, the two stellar cores lose orbital energy, which is deposited within the envelope and leads to a shortening of the binary period (e.g. Ricker \& Taam \cite{ricker08}). Eventually the common envelope is ejected and a close binary system is formed, which contains a core helium-burning sdB and a main-sequence companion. If the second star reaches the red giant branch, another common envelope phase is possible and can lead to a close binary with a white dwarf companion and an sdB. 

EGB\,5 was discovered to be the blue central star of a faint elliptical planetary nebula (PN\,G\,211.9+22.6) by Ellis et al. (\cite{ellis84}). M\'endez et al. (\cite{mendez88a}) derived the atmospheric parameters of this star by fitting model spectra. The resulting parameters $T_{\rm eff}=42\,000\pm5000\,{\rm K}$ and $\log{g}=5.8\pm0.2$ were neither consistent with canonical post-asymptotic giant branch (post-AGB) evolutionary tracks nor the parameters of all other known central stars of planetary nebulae (CSPN, Drilling \& Sch\"onberner \cite{drilling85}). M\'endez et al. (\cite{mendez88a}) recognized that these parameters are typical of a hot subdwarf star rather than a post-AGB object and speculated that this is a close binary that experienced non-conservative mass exchange during the first giant phase. 

Lisker et al. (\cite{lisker05}) derived more accurate atmospheric parameters of EGB\,5  from high resolution spectra obtained in the course of the SPY survey, which are perfectly consistent with a core helium-burning sdB star ($T_{\rm eff}=34\,000\pm400\,{\rm K}$, $\log{g}=5.85\pm0.05$, $\log{y}=-2.77\pm0.04$). 

Only one CSPN candidate with similar parameters is known so far (PHL\,932, M\'endez et al. \cite{mendez88b}; Napiwotzki \cite{napiwotzki99}; Lisker et al. \cite{lisker05}). Whether this star has a close companion remains disputed, because several searches for RV variations yielded inconsistent results (Wade \cite{wade01}; de Marco et al. \cite{demarco04}; Af\c{s}ar \& Bond \cite{afsar05}). It is therefore unclear whether this star experienced a common envelope phase. Furthermore, Frew et al. (\cite{frew10}) convincingly demonstrated that the nebula around PHL\,932 is not a planetary nebula, but rather a Str\"omgren sphere.

Here we report the discovery of a close companion to EGB\,5.

\begin{figure}[t!]
	\centering
	\resizebox{\hsize}{!}{\includegraphics{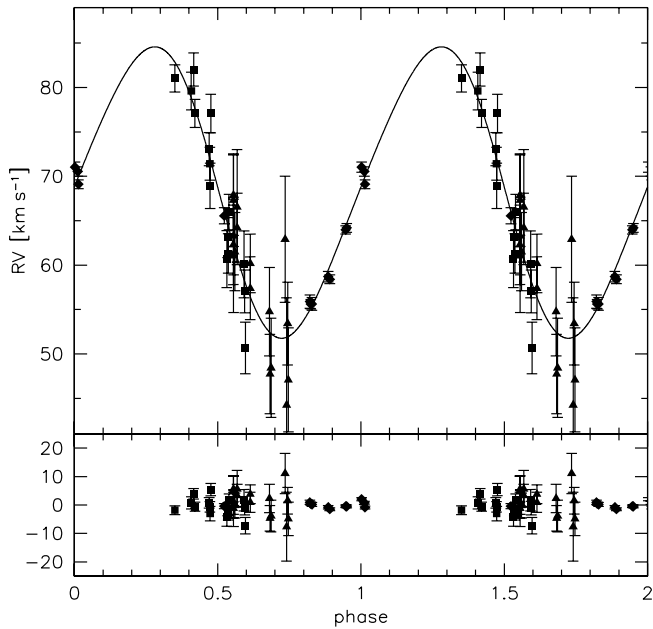}}
	\caption{Radial velocities of the subdwarf primary measured from spectra obtained with UVES (rectangles), ISIS (upward triangles), and TWIN (diamonds) plotted against orbital phase ($P=16.537\,{\rm d}$). The residuals are plotted below.}
	\label{rv}
\end{figure}

\section{Orbital parameters}

EGB\,5 was observed twice in the course of the SPY project with the high resolution echelle spectrograph UVES at the ESO-VLT. Follow-up high resolution spectra were obtained with UVES, medium resolution spectra were taken with the ISIS spectrograph at the WHT, and the TWIN spectrograph mounted at CAHA\,3.5m telescope. The RVs were measured by fitting a set of mathematical functions (Gaussians, Lorentzians, and polynoms) to the hydrogen Balmer and helium lines using the FITSB2 routine (Napiwotzki et al. \cite{napiwotzki04b}). Errors were calculated with a bootstrapping algorithm. The RV measurements are given in Table~\ref{RVs}.

\begin{figure}[t!]
	\centering
	\resizebox{\hsize}{!}{\includegraphics{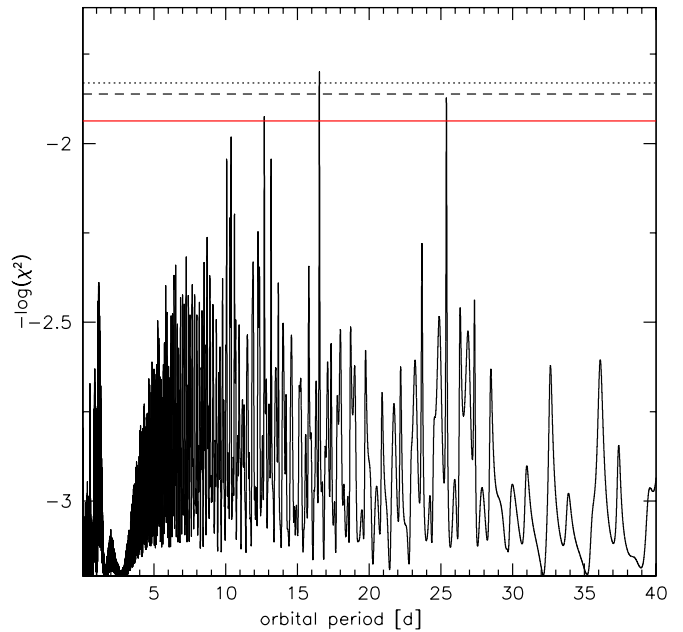}}
	\caption{In this power spectrum $-\log\chi^{2}$ of the best sine fits is plotted against the orbital periods. The dotted horizontal line marks the $1\sigma$ confidence limit, the dashed the $3\sigma$, and the solid line the $6\sigma$ limit.}
	\label{power}
\end{figure}

To determine the orbital parameters and estimate the significance of the solution, sine curves were fitted to 43 RV data points using a $\chi^{2}$-minimising method (SVD). The $\chi^2$ against orbital period is given in Fig.~\ref{power}. Two peaks are present at $P\simeq16.53\,{\rm d}$ and $P\simeq25.39\,{\rm d}$ with the former one being the most probable solution. We performed Monte Carlo simulations ($10\,000$ iterations) to estimate the significance of the solution. For each simulation, a randomised set of RVs was drawn from Gaussian distributions with central values and widths corresponding to the RV measurements and the analysis repeated. In $82\%$ of the trials, the most likely solution was fitted, the next best solution being chosen in $12\%$ of the iterations. 

Since the RV variation in EGB\,5 is very small, we used FITSB2 to perform a simultaneous fit of Keplerian orbits to all 169 lines in the 43 spectra covering different orbital phases, i.e., all available information is combined into the parameter determination procedure. Motivated by the discovery of eccentric orbits in close sdB binaries (Edelmann et al. \cite{edelmann05}; Napiwotzki et al. in prep.), we did not restrict our fitting to circular orbits. The best-fit orbital solution is shown in Table~\ref{tab:par}. The errors were determined by bootstrapping and the reduced $\chi^{2}$ of the best-fit solution is $0.97$. For comparison, a sine curve was fitted to the single RV points in the way described in Geier et al. (\cite{geier10c}), and our derived orbital parameters ($P=16.532\,{\rm d}$, $\gamma=68.8\,{\rm km\,s^{-1}}$, $K=16.1\,{\rm km\,s^{-1}}$) are consistent with the results given in Table~\ref{tab:par}. 

We are reluctant to claim orbital eccentricity in the case of EGB\,5 despite the formal eccentricity being found to be $0.098\pm0.048$. An F-test indicates a borderline significance of $87\%$ compared to a circular solution. Systematic errors introduced by combining datasets taken with different instruments can easily mimic small eccentricities. High-resolution, time-resolved spectroscopy with a better coverage of the whole orbit would be necessary to solve this issue. Since the formal eccentricity is low, it does not significantly affect the derived minimum companion mass.

\section{Atmospheric parameters \label{sec:ana}}
 
The atmospheric parameters of EGB\,5 were determined by fitting LTE models with enhanced UV metal line blanketing to the coadded high S/N UVES spectrum. Using models with solar metallicity, the Balmer lines and the He\,{\sc ii} line at $4686\,{\rm \AA}$ cannot be fitted simultaneously. We choose ten times solar metallicity models to mimic strong UV line blanketing. This is motivated by the strong enrichment of heavy elements produced by radiative levitation in sdB stars with similar parameters (O'Toole \& Heber \cite{otoole06}; Geier et al. \cite{geier07}). The atmospheric parameters (see Table~\ref{tab:par}) are similar to the ones derived by Lisker et al. (\cite{lisker05}).

\begin{table}
\caption{Radial velocities of EGB\,5}
\label{RVs}
\begin{center}
\begin{tabular}{lll}
\hline
\noalign{\smallskip}
mid-HJD-2450000 & RV [${\rm km\,s^{-1}}$] & Instrument\\
\noalign{\smallskip}
\hline
\noalign{\smallskip}
$2008.58634$ & $69.1\pm0.5$ & UVES \\
$2033.54169$ & $65.6\pm0.9$ & \\
$2716.50684$ & $55.9\pm0.7$ & \\
$2716.55553$ & $55.6\pm0.5$ & \\
$2716.63020$ & $55.6\pm0.7$ & \\ 
$2717.57133$ & $58.7\pm0.6$ & \\
$2717.65759$ & $58.4\pm0.5$ & \\ 
$2718.57307$ & $64.0\pm0.5$ & \\
$2718.63521$ & $64.2\pm0.5$ & \\
$2719.49370$ & $71.0\pm0.6$ & \\
$2719.65305$ & $70.6\pm0.5$ & \\
\noalign{\smallskip}
\hline
\noalign{\smallskip}
$2328.35974$ & $81.0\pm1.5$ &  TWIN\\
$2329.30317$ & $79.6\pm2.1$ & \\ 
$2329.44255$ & $82.0\pm1.9$ & \\ 
$2329.52035$ & $77.1\pm1.6$ & \\  
$2330.33790$ & $73.1\pm1.9$ & \\ 
$2330.38145$ & $71.4\pm1.9$ & \\  
$2330.39448$ & $68.9\pm2.6$ & \\ 
$2330.43186$ & $77.1\pm2.2$ & \\ 
$2331.35629$ & $60.7\pm3.2$ & \\ 
$2331.40692$ & $63.1\pm2.3$ & \\ 
$2331.43920$ & $61.3\pm2.2$ & \\ 
$2331.48586$ & $66.0\pm2.0$ & \\ 
$2332.34960$ & $60.1\pm3.8$ & \\
$2332.38591$ & $57.1\pm2.4$ & \\ 
$2332.42981$ & $50.7\pm2.9$ & \\
\noalign{\smallskip}
\hline
\noalign{\smallskip}
$2662.47992$ & $62.3\pm4.5$ & ISIS\\
$2662.48416$ & $67.9\pm4.5$ & \\
$2662.48781$ & $67.3\pm5.2$ & \\ 
$2662.49147$ & $61.1\pm6.4$ & \\
$2662.49513$ & $63.3\pm4.4$ & \\
$2662.57002$ & $61.5\pm4.4$ & \\
$2662.69661$ & $66.5\pm6.5$ & \\
$2662.70027$ & $64.2\pm3.9$ & \\
$2663.45598$ & $57.4\pm3.5$ & \\
$2663.45964$ & $60.2\pm3.3$ & \\ 
$2664.56672$ & $54.8\pm5.0$ & \\
$2664.59451$ & $47.7\pm4.5$ & \\
$2664.65670$ & $48.4\pm5.6$ & \\
$2665.46532$ & $62.9\pm7.1$ & \\
$2665.54901$ & $44.3\pm12.1$ & \\
$2665.61243$ & $53.4\pm4.7$ & \\
$2665.65533$ & $47.1\pm5.9$ & \\
\noalign{\smallskip}
\hline

\end{tabular}
\end{center}
\end{table}

\begin{table}[h!]
\caption{Binary parameters of EGB\,5.} 
\label{tab:par}
\begin{center}
\begin{tabular}{ll}
        \hline
        \noalign{\smallskip}
        Orbital parameters & \\
        \noalign{\smallskip}
        \hline
        \noalign{\smallskip}
        $T_{\rm 0}$ [HJD]    & $2452719.457\pm0.055$ \\
        $P$                  & $16.537\pm0.003\,{\rm d}$ \\
        $\gamma$             & $68.5\pm0.7\,{\rm km\,s^{-1}}$ \\
        $K$                  & $16.1\pm0.8\,{\rm km\,s^{-1}}$ \\
        $e$                  & $0.098\pm0.048$ \\
        $\Omega$             & $102\pm59^{\rm \circ}$ \\
        $f(M)$               & $0.0072\pm0.0011\,M_{\rm \odot}$ \\
        \noalign{\smallskip}
        \hline
        \noalign{\smallskip}
        Atmospheric parameters & \\
        \noalign{\smallskip}
        \hline
        \noalign{\smallskip}
        $T_{\rm eff}$        & $34500\pm500\,{\rm K}$ \\
        $\log{g}$             & $5.85\pm0.05$ \\
        $\log{y}$             & $-2.9\pm0.09$ \\
        \noalign{\smallskip}
        \hline
        \noalign{\smallskip}
        Derived binary parameters & \\
        \noalign{\smallskip}
        \hline
        \noalign{\smallskip}
        $M_{\rm 1}$ (adopted) & $0.47\,M_{\rm \odot}$ \\
        $R_{\rm 1}$           & $0.13\,R_{\rm \odot}$ \\
        $M_{\rm 2,min}$       & $0.14\,M_{\rm \odot}$ \\
        $a_{\rm min}$         & $23\,R_{\rm \odot}$ \\
        \noalign{\smallskip}
        \hline
\end{tabular}
\end{center}
\end{table}

\section{Nature of the unseen companion}

Adopting the canonical sdB mass of $0.47\,M_{\rm \odot}$, the minimum mass of the unseen companion ($0.14\,M_{\rm \odot}$) is consistent with either a late main-sequence star of spectral type M or a low-mass white dwarf. Since no spectral features of the companion are visible in the optical spectra, a main-sequence companion with a mass higher than $\simeq0.45\,M_{\rm \odot}$ can be excluded (Lisker et al. \cite{lisker05}). A white dwarf companion of similar or more mass would require the binary inclination to be lower than $24^{\rm \circ}$. Assuming randomly distributed inclinations, the probability for such a low inclination in the case of EGB\,5 is less than $9\%$.

Owing to its long orbital period it is rather unlikely that the system is eclipsing or that it shows other detectable  features (e.g. reflection effects) in its light curve. It is therefore not possible to constrain the nature of the companion further. The low mass indicates that it is either a late M dwarf or a low mass WD with a He core.

For the most likely companion mass range of $0.14-0.45\,M_{\rm \odot}$, the separation between sdB and companion is constrained to be $23-27\,R_{\rm \odot}$. All relevant measurements and parameters of the EGB\,5 system are summarized in Table~\ref{tab:par}.

\section{Discussion}

EGB\,5 has the second longest period of all known sdBs in close binary systems. Morales-Rueda et al. (\cite{morales03}) discovered three binaries with periods near or exceeding $10\,{\rm d}$ and unseen companions (PG\,0850$+$170, $27.815\,{\rm d}$; PG\,1619$+$522, $15.3578\,{\rm d}$; PG\,1110$+$294, $9.4152\,{\rm d}$). In all these cases, the derived minimum companion masses exceed $0.45\,M_{\rm \odot}$ consistent with WD companions. The low minimum companion mass of EGB\,5 is therefore rather unusual. 

Close binaries with sdB primaries and M dwarf companions have been discovered by means of variations in their light curves caused by light originating in the heated surface of the cool companion and often accompanied by eclipses (see e.g. For et al. \cite{for10}; \O stensen et al. \cite{oestensen10}). The orbital periods of these binaries are very short ($\simeq0.1-0.3\,{\rm d}$), because these so-called reflection effects can only be detected in such cases. The longest period systems, where a reflection effect was detected, is the binary JL\,82 ($0.7371\,{\rm d}$, Edelmann et al. \cite{edelmann05}; Koen  \cite{koen09}; Geier et al. \cite{geier10b}). 

With an orbital period of $16.537$ days, EGB\,5 would have by far the longest period of all sdB+dM binaries known. The M-dwarf companion must have been engulfed by the giant progenitor of the sdB star and the common envelope ejected. According to theoretical models, the change in orbital energy must be at least as large as the binding energy of the giant (see Paczynski \cite{paczynski76}; Han et al. \cite{han02,han03}). Detailed models of giants at the tip 
of the red-giant branch that will form sdB stars after the common envelope are given in Hu et al. (\cite{hu07}). For a $1\,M_{\rm \odot}$ progenitor, even a low mass companion of $0.14\,M_{\rm \odot}$  in principle could eject the envelope, at least if the thermal energy of the envelope can be used. For more massive companions, the envelope could even be ejected without the thermal energy. However, in any case it implies that the efficiency of the common envelope ($\alpha$) is high. Assuming that the efficiency is a universal parameter, this means that the much closer HW\,Vir binaries should have formed from more massive sdB progenitors and thus should have lower masses (see Hu et al. \cite{hu07}, Fig.~3) than EGB\,5. In contrast, the masses of most sdB primaries in HW\,Vir systems constrained by observations seem to be close to the canonical value. However, we point out that reliable mass determinations in such systems are still hampered by severe issues (see e.g. For et al. \cite{for10}; \O stensen et al. \cite{oestensen10}). de Marco et al. (\cite{demarco11}) suggested that low mass companions may have higher common envelope efficiencies, possibly because of their longer in-spiral timescales.

While close binary sdB+dM systems of HW\,Vir type are easy to find from their characteristic light curves, it is much harder to discover these systems in wider orbits and at lower inclination. Were the companion of EGB\,5 to be an M dwarf, these systems might be very common and that $\simeq10\%$ of all known sdB binaries are of HW\,Vir type just a selection effect. Barlow et al. (\cite{barlow10}) found a sinusoidal variation in the O-C-diagram of the pulsating sdB CS\,1246, which is most likely caused by a low mass companion in a $14.1\,{\rm d}$ orbit very similar to EGB\,5. 

The period distribution of sdBs with confirmed WD companions is much wider (Morales-Rueda et al. \cite{morales03}; Geier et al. \cite{geier10b}), but the combination of low mass and long period make EGB\,5 peculiar. All known detached double-degenerate systems with very low-mass WDs have much shorter periods (Steinfadt et al. \cite{steinfadt10} and references therein), which may again be caused by selection effects. 

Furthermore, EGB\,5 may be the first very young post-CE system, where the ejected envelope is still visible (Mendez et al. \cite{mendez88a}) and may provide direct evidence of this sdB formation channel. This would make this binary a unique system to study this short and poorly understood phase of binary evolution. 

PHL\,932 has been regarded as a similar object. However, evidence that it has a close companion is not compelling. In addition, Frew et al. (\cite{frew10}) showed that the nebulous structure around PHL\,932 is not a planetary nebula, but rather a Str\"omgren sphere created by the hot sdB in the surrounding interstellar medium. Since EGB\,5 is located at relatively low Galactic latitude ($+22.6^{\rm \circ}$) this possibility should be seriously investigated.

\begin{acknowledgements}

S.~G. is supported by the Deutsche Forschungs\-gemeinschaft under grant HE1356/49-1. 

\end{acknowledgements}

\end{document}